\documentclass[aps,prl,twocolumn,groupedaddress,floats,showpacs]{revtex4}
\usepackage{latexsym}
\usepackage{dcolumn}
\usepackage[dvips]{graphicx}
\usepackage{amssymb}
\usepackage{graphics}
\usepackage{amsmath}
\usepackage{epsf}

\newcommand{\vk}{{\bf k} }
\newcommand{\vm}{\hat{{\bf m}} }
\newcommand{\ve}{\hat{{\bf e}} }
\newcommand{\vq}{{\bf q} }

\newcommand{\up}{\uparrow}
\newcommand{\down}{\downarrow}

\begin{document}
 
\title{Mesoscopic anisotropic magnetoconductance fluctuations in ferromagnets}

\author{Shaffique Adam, Markus Kindermann, Saar Rahav, and Piet W. Brouwer}

\affiliation{Laboratory of Atomic and Solid State Physics,
Cornell University, Ithaca, NY 14853-2501}
\date{\today}

\begin{abstract}
The conductance of a ferromagnetic particle depends on the relative
orientation of the magnetization with respect to the direction of
current flow. This phenomenon is known as ``anisotropic
magnetoresistance''. Quantum interference leads to an additional,
random dependence of the conductance on the magnetization
direction. These ``anisotropic magnetoresistance fluctuations'' are
caused by spin-orbit scattering, which couples the electron motion to
the exchange field in the ferromagnet. We report a calculation of the
dependence of the conductance autocorrelation function on the rotation
angle of the magnetization direction.   
\end{abstract}


\maketitle

One hallmark of phase-coherent transport is the phenomenon of
``universal conductance fluctuations'', random, but reproducible 
variations in a sample's conductance as a function of the applied
magnetic field or the Fermi energy
\cite{kn:lee1985b,kn:lee1987,kn:altshuler1985b,kn:altshuler1986b,kn:altshuler1986}.
The magnitude of the conductance fluctuations is of order unity,
in units of the conductance quantum $e^2/h$, and does not depend on 
specific sample properties, such as the impurity concentration, the
meterial, shape, or method of preparation.

Recently there has been both theoretical and experimental interest
in mesoscopic transport in itinerant ferromagnets. The experimental
interest stems from the ability to fabricate ferromagnetic conductors
small enough that transport through the magnet is predominantly
coherent \cite{kn:lee2004,kn:wei2005}. The theoretical interest is motivated by
the rich variety of ways through which random impurity scattering can 
affect the properties of an itinerant ferromagnet. Theoretical
predictions exist for the effect 
of domain walls on weak localization and conductance fluctuations 
\cite{kn:tatara1997,kn:lyandageller1998} as well as for the combined effect of spin-orbit 
interaction and impurity scattering on weak localization
\cite{kn:dugaev2001} and magnetic anisotropy \cite{kn:brouwer2005b}. Although
disordered ferromagnetic conductors display different phenomena than 
their normal-metal counterparts, the theoretical framework to describe them
is rather similar. Indeed, the
methods of diagrammatic perturbation theory developed for electron
transport in disordered metals can be applied to ferromagnets by
modifying the single particle Hamiltonian taking into account the
exchange field and/or spin-orbit interactions.  

In this Letter, we address the mesoscopic contribution to a
ferromagnet's anisotropic magnetoresistance. Anisotropic
magnetoresistance is the phenomenon that a magnet's resistance depends
on the orientation of the magnetization resulting from a
combination of spin-orbit coupling and orbital magnetic effects
\cite{kn:ohandley2000}. For a single domain magnet, the resistance is a
smooth function of the magnetization direction. The mesoscopic effect
described here consists of an additional and faster random dependence
on the magnetization direction that is different for each sample, but
reproducible for a given sample. This situation is not very different
from the case of standard universal conductance fluctuations in a
normal metal, where the random magnetic-field dependent fluctuations
are superimposed on a systematic magnetoconductance.

There are two possible mechanisms through which the magnetization
direction can affect the interference correction to the
conductance. First, a change of the magnetization direction causes a
change of the internal magnetic field, which directly affects the
orbital motion of the electrons via a change of Aharonov-Bohm
phases. Second, a change of the magnetization direction causes a
change of the exchange field, which affects the motion of the electrons
via spin-orbit scattering. The first effect would be dominant if the
magnetic flux through the cross-section of a phase coherent volume is 
of the order of the flux quantum. For many
magnetic materials, the phase coherent lengths can be small and this
effect can be neglected (see discussion in
Ref.~\onlinecite{kn:dugaev2001}). In what follows, we assume that this
condition holds, and that the second effect dominates the mesoscopic
anisotropic magnetoresistance. For the same reason, we ignore any
effect of an applied magnetic field used to change the magnetization 
direction.

We consider an ensemble of ferromagnetic particles, each with a
different configuration of impurities and calculate the conductance
autocorrelation function
\begin{equation}
{\cal C} (\theta) = \langle G(\vm) G(\vm') \rangle - \langle G(\vm)
\rangle^2,
  \label{eq:C}
\end{equation}
where $\theta$ is the angle between the magnetization directions $\vm$
and $\vm'$ and the brackets $\langle \ldots \rangle$ denote the ensemble
average. The vectors $\vm$ and $\vm'$ are defined to have unit length.
The Hamiltonian for a ferromagnet with spin-orbit scattering is
\begin{equation}
  {\cal H}_{\alpha \beta} = \left( \frac{p^2}{2m} - \mu \right)
  \delta_{\alpha \beta}
  - E_{\rm Z} \sigma^z_{\alpha\beta} + {\cal V}_{\alpha\beta}
\label{Eq:Hamiltonian}
\end{equation}  
where $\alpha$ and $\beta$ are spin indices, $\sigma_z$ the
Pauli matrix, the magnetization direction $\vm$ is taken as the spin
quantization axis, and $E_{\rm Z} = \mu_{\rm B} 
B_{\rm ex}$ is the Zeeman energy 
corresponding to the exchange field $B_{\rm ex}$. We perform the
ensemble average at a fixed chemical potential $\mu$ and exchange
field $B_{\rm ex}$, rather than at self-consistently determined $\mu$
and $B_{\rm ex}$. Although the omission of the self-consistency
conditions is known to affect averaged quantities, it is believed not
to affect fluctuations 
\cite{kn:bouchiat1989,kn:cheung1989,kn:altshuler1991}.

The random potential ${\cal V}$ in Eq.\ (\ref{Eq:Hamiltonian})
describes the effect of elastic impurity scattering and spin-orbit
scattering, respectively. Its Fourier transform is
\begin{eqnarray}
  \label{eq:Vso}
  {\cal V}_{\alpha \vk, \beta \vk'} &=&
  V_{\vk - \vk'} \\ && \mbox{}
  -i V^{\rm so}_{\vk - \vk'} ((\vk' \times \vk) \cdot
  (\vm \sigma^{z} + \ve_1 \sigma^x
  + \ve_2 \sigma^{y})_{\alpha \beta}, \nonumber
\end{eqnarray}
where $\ve_1$ and $\ve_2$ are unit vectors perpendicular to 
each other and to $\vm$ such that $\ve_1 \times \ve_2 = \vm$.  
The random potentials $V$ and $V^{\rm so}$ 
are assumed to be uncorrelated and Gaussian white noise, with r.m.s.\ 
strength $v$ and $v^{\rm so}$, respectively,
\begin{equation}
  \langle V_{\vq} V_{\vq'} \rangle =
  v^2 \delta(\vq-\vq'),\ \
  \langle V^{\rm so}_{\vq} V^{\rm so}_{\vq'} \rangle =
  v_{\rm so}^2 \delta(\vq-\vq').
\end{equation}
In the leading order Born approximation,
the scattering time $\tau_{\alpha}$ for spin-independent impurity 
scattering of electrons with spin $\alpha$ is given by
\begin{equation}
  \frac{1}{2 \pi \nu_{\up}
  \tau_\up} = v^2,\ \
  \frac{1}{2 \pi \nu_{\down} \tau_{\down}} = v^2,
  \label{eq:tau}
\end{equation}
where $\nu_{\alpha}$ is the density of states of electrons with spin
$\alpha$. Similarly, for spin-conserving and spin-flip 
scattering off $V^{\rm so}$, one has the mean free times
\begin{eqnarray}
  \frac{1}{2 \pi  \nu_{\up} \tau_{\up \parallel}} &=&
  \frac{2}{9} v_{\rm so}^2 k_{{\rm {\rm F}}
  \up}^4,\ \ \frac{1}{2 \pi  \nu_{\down} \tau_{\down \parallel}} =
  \frac{2}{9} v_{\rm so}^2 k_{{\rm F}
  \down}^4, \nonumber \\
  \frac{1}{2 \pi \nu_{\down} \tau_{\up\perp}} &=& 
  \frac{1}{2 \pi \nu_{\up} \tau_{\down\perp}} =
  \frac{2}{9} v_{\rm so}^2 k_{{\rm F} \up}^2 k_{{\rm F} \down}^2,
  \label{eq:taus}
\end{eqnarray}
respectively, where $k_{{\rm F} \alpha}$ is the Fermi wavevector for spin 
$\alpha$ electrons. In a realistic ferromagnet, the kinetic energy and
the random potential will not have the simple form assumed in our
calculation, which implies that the relationships between the
scattering times implied by Eqs.\ (\ref{eq:tau}) and (\ref{eq:taus})
need not hold. Although we use the simple model described above to set
up our calculation and to define the scattering times, these are then 
considered independent for the rest of the calculation [except for the
equality in the second line of Eq.\ (\ref{eq:taus}), which follows
from detailed balance].

Throughout the calculation, we assume that $\tau \ll 
\tau_{\parallel}$, $\tau_{\perp}$. This implies that all Green functions
appearing in intermediate phases of the calculation can be averaged
over all directions of the momentum. We also assume
that phase coherence is preserved over the entire sample.
In a sample with size $L$ larger than the phase coherence 
length $L_\phi$, our answer would be modified as ${\cal C}(\theta,
L) \sim
{\cal C} (\theta, L_\phi) (L_\phi/L)$.  In this case, the angle
over which the conductance typically fluctuates is then determined by
$L_\phi$ instead of $L$. 
   
\begin{figure}
\bigskip
\epsfxsize=0.8\hsize
\hspace{0.0\hsize}
\epsffile{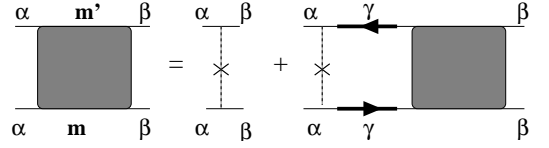}
\caption{\label{Fig:Dyson} Dyson Equation for Diffuson ladder.  The dotted line
indicates a scattering event.}
\end{figure}

We now describe the details of our calculation. For the 
retarded Green function 
${\cal G}^{\rm R}$, averaged over the random potential and over all
directions of the momentum, we find
\begin{equation}
  \langle{{\cal G}^{\rm R}_{\alpha}(\omega, k, \vm)}\rangle^{-1} =
  \omega - \varepsilon_{\alpha}(k)
  + \frac{i}{2 \tau_{\alpha}}
  + \frac{i}{2 \tau_{\alpha\parallel}}
  + \frac{i}{\tau_{\alpha\perp}},
  \label{eq:G}
\end{equation}
where 
$\varepsilon_{\alpha}(k) = \hbar^2 k^2/2m - \mu - E_{\rm Z} 
\sigma^{z}_{\alpha\alpha}$ is the energy of an electron with spin
$\alpha$ and momentum $\hbar \vk$. In order to calculate the
conductance autocorrelation function (\ref{eq:C}), we need to consider
the Diffuson and Cooperon propagators of diagrammatic perturbation
theory. Again, in view of the inequality $\tau \ll \tau_{\parallel}$,
$\tau_{\perp}$, we only need Diffuson and Cooperon propagators averaged
over all momentum directions.
Since the Cooperon and Diffuson propagators are related by time
reversal,
\begin{equation}
  C(\omega, \vq, \theta) = D(\omega, \vq, \pi - \theta),
  \label{eq:DC}
\end{equation}
it will be sufficient to calculate the Diffuson only. 

The
Diffuson propagator is defined by the ladder diagrams 
shown in Fig.\ \ref{Fig:Dyson}. The solid arrows in Fig.\
\ref{Fig:Dyson} denote the impurity-averaged Green functions
(\ref{eq:G}). The two legs of the ladder refer to
the two magnetization directions $\vm$ and $\vm'$. For both
magnetization directions we use the convention that the magnetization
direction is the spin quantization axis. This is the natural choice
for ferromagnets: Since $E_{\rm Z}
\tau \gg 1$ in a typical ferromagnet, with this convention only 
ladder diagrams for which 
the spin indices of retarded and advanced Green functions are pairwise
equal at 
all times need to be considered; contributions with different spin index
for retarded and advanced Green functions dephase within a mean free
time and do not contribute to the Diffuson propagator. One should note,
however, that this convention implies that the directions
of ``spin up'' and ``spin down'' in the upper and lower legs of the
ladder correspond to different physical directions if $\vm \neq \vm'$.

Summing the ladder
diagrams of Fig.\ \ref{Fig:Dyson}, we then find that the Diffuson
obeys the $2 \times 2$ matrix equation
\begin{equation}
  \sum_{\gamma=\up,\down}
  K_{\alpha \gamma} D(\omega,\vq,\theta)_{\gamma \beta} = \delta_{\alpha \beta}
  \frac{1}{2 \pi \nu_{\alpha} \tau_{\alpha}}.
\end{equation}
Here $K$ is a $2 \times 2$ matrix, with diagonal elements given by
\begin{eqnarray}
  \hat{K}_{\alpha \alpha} &=&  \tau_\alpha 
  \left[ D_\alpha q^2 + i \omega +
  \frac{2}{\tau_{\alpha\perp}} + \frac{1- \cos
  \theta}{\tau_{\alpha\parallel}}
  \right],
\end{eqnarray}
where $D_\alpha = v_{{\rm F}\alpha}^2 \tau_\alpha/3$ is the diffusion
constant. The off-diagonal matrix elements contain a phase factor that 
depends on the precise choice of coordinate axes perpendicular to
$\vm$ and $\vm'$, cf.\ 
Eq.\ (\ref{eq:Vso}). In all final expressions, the off-diagonal
elements of $K$ only enter 
through their product, which is independent of this choice,
\begin{equation}
  K_{\up \down} K_{\down \up} = \frac{\tau_{\up}
    \tau_{\down}}{\tau_{\up\perp} \tau_{\down\perp}}
  (1 + \cos \theta)^2.
\end{equation}
Once the Diffuson is known, the Cooperon is calculated via Eq.\
(\ref{eq:DC}).
For the special case $\theta=0$, the result for $C$ was previously
obtained by Dugaev {\em et al.}\cite{kn:dugaev2001}. 

\begin{figure}
\bigskip
\epsfxsize=0.8\hsize
\hspace{0.00\hsize}
\epsffile{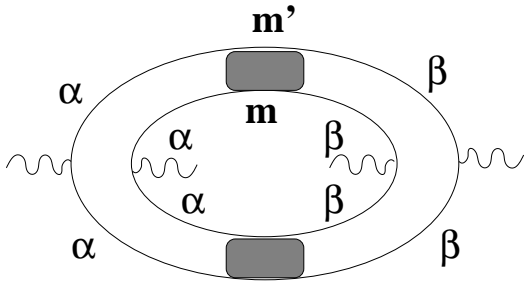}
\caption{\label{Fig:UCF} Leading diagrams for conductance correlator.
The wavy lines represent the current vertex $J$ and the shaded box 
represents either the Cooperon or Diffuson propagator.}
\end{figure}

We can now proceed to calculate the conductance correlation function
${\cal C}(\theta)$. We are interested in the conductance
correlations at zero temperature, which 
allows us to set $\omega=0$ in our expressions for the Diffuson and 
Cooperon propagators. We consider a coherent rectangular sample
with sides $L_x, L_y$ and $L_z$, with a current in the $z$ 
direction. Following 
Refs.~\onlinecite{kn:altshuler1986,kn:lee1987}, we then find that the conductance 
autocorrelation function 
is given by
\begin{equation}
  {\cal C}(\theta) = 
  \frac{3 e^4}{2 L_z^4 h^2} \sum_{\vq} \mbox{tr}\, [ J D(\vq) J
  D(\vq) + J C(\vq) J
  C(\vq) ],
  \label{eq:Corr}
\end{equation}
where the current vertex reads
\begin{eqnarray}
  J_{\alpha \beta} = \frac{4 \pi}{3} v_{{\rm F} \alpha}^2
  \tau_{\alpha}^3 \delta_{\alpha \beta}
\end{eqnarray}
and the vector $\vq$ is summed over the values $q_x = \pi n_x/L_x$,
$q_y = \pi n_y/L_y$, and $q_z = \pi n_z/L_z$ with $n_{x}$, $n_{y} =
0,1,2,\ldots$ and $n_z = 1,2,\ldots$. Without the prefactor $3/2$, 
Eq.\ (\ref{eq:Corr}) is precisely the contribution from the diagram
shown in Fig.\ \ref{Fig:UCF}. The factor $3/2$ in front
of Eq.\ (\ref{eq:Corr}) accounts for other diagrams that
contribute to the conductance fluctuations, whose net contribution is
$1/2$ times that of the diagram of Fig.\ \ref{Fig:UCF}
\cite{kn:lee1987,kn:altshuler1986}.  Substituting our results for the Diffuson and Cooperon propagators, we find
\begin{widetext}
\begin{eqnarray}
{\cal C}(\theta) = \frac{6e^4}{\pi^4 h^2} \sum_{\vq} \sum_{\pm}
  \left[
  \frac{1}{((L_z q/\pi)^2 + a_{\pm}(\theta))^2} + 
  \frac{1}{((L_z q/\pi)^2 + a_{\pm}(\pi-\theta))^2}
  \right],
  \label{eq:TrLam}
\end{eqnarray}
where 
\begin{eqnarray}
  a_{\pm}(\theta) &=&
  \frac{1}{\tau_{\up\perp} E_{\up}} +
  \frac{1}{\tau_{\down\perp} E_{\down}} +
  \frac{\tau_{\up\parallel} E_{\up} + \tau_{\down\parallel} E_{\down}}
  {2 \tau_{\up\parallel} \tau_{\down\parallel} E_{\up} E_{\down}}
  (1 - \cos \theta))
  \nonumber \\ && \mbox{}
  \pm
  \sqrt{ \frac{(1 + \cos \theta)^2}{\tau_{\up\perp} \tau_{\down\perp}
  E_{\up}E_{\down}}
  +
  \left[\frac{1}{\tau_{\up\perp} E_{\up}} -
  \frac{1}{\tau_{\down\perp} E_{\down}} -
  \frac{\tau_{\up\parallel} E_{\up} - \tau_{\down\parallel} E_{\down}}
  {2 \tau_{\up\parallel} \tau_{\down\parallel} E_{\up} E_{\down}}
  (1 - \cos \theta) \right]^2}
\label{eq:main}
\end{eqnarray}
\end{widetext}
and $E_{\alpha} = D_{\alpha} (\pi/L_z)^2$ is the Thouless energy for
spin $\alpha$. Note that the parameter that governs the importance of
spin-orbit scattering is the product $\tau_{\alpha \perp} E_{\alpha}$
or $\tau_{\alpha \parallel} E_{\alpha}$, which is the ratio of the
spin-orbit time and the Thouless time, which is 
the time to diffuse through the
sample.

The expression for $a_{\pm}(\theta)$ simplifies in two limiting
cases. If $\theta=0$, one has $a_+ = 2/\tau_{\up\perp} E_\uparrow + 
2/ \tau_{\down\perp} E_\downarrow$ and $a_{-} = 0$, showing the
presence of universal conductance fluctuations in a ferromagnet. 
The corresponding eigenvalues for the Cooperon
contribution are found by setting $\theta=\pi$, $a_+(\pi) =
2/\tau_{\up\perp} E_{\up} + 2/\tau_{\up\parallel} E_{\up}$ and
$a_-(\pi) = 2/\tau_{\down\perp} E_{\down} + 2/\tau_{\down\parallel}
E_{\down}$. Another simple limit is that of a half-metal, a
ferromagnet with vanishing density of states for the minority
spins. For a half metal, the only relevant time and energy scales are
the scattering time $\tau_{\up\parallel}$ for spin-preserving
spin-orbit scattering of majority electrons and the majority
electron Thouless energy $E_{\up}$. One then finds that only one
root $a_{\pm}$ is relevant,
$a(\theta) = (1 - \cos \theta)/\tau_{\up\parallel} E_{\up}$.

The sum over wavevectors in Eq.~\ref{eq:TrLam} can be performed
analytically for a quasi one-dimensional sample. Setting $n_x = n_y =
0$ in the summation, one finds
\begin{eqnarray}
  {\cal C}(\theta) &=&
  \sum_{\pm} \left[ F(\pi \sqrt{a_{\pm}(\theta)}) +
  F(\pi \sqrt{a_{\pm}(\pi-\theta)})\right],
  \label{eq:Cresult}
\end{eqnarray}
where $F(x) = 3e^4(-2 + x \coth x + x^2 \sinh^{-2} x)/2 x^4 h^2$.
Note that for $\theta = 0$,
Eq.\ (\ref{eq:Cresult}) reproduces the known results $\mbox{var}\, G =
(e^2/h)^2 (1/15)$ for strong spin-orbit scattering
and $\mbox{var}\, G = (e^2/h)^2 (4/15)$ for weak spin orbit scattering.
\begin{figure}
\bigskip
\epsfxsize=0.8\hsize
\hspace{0.05\hsize}
\epsffile{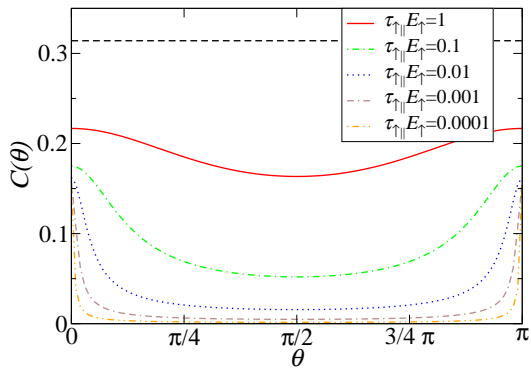}
\caption{\label{Fig:Numerics} 
The correlation function of the conductance at different directions of the 
magnetization, for various strengths of the spin orbit
scattering. Results
shown here are for a half metal with cubic geometry.}
\end{figure}
For quasi 2D and 3D samples ${\cal C}(\theta)$ can be computed
numerically.  The dependence on the spin-orbit scattering is
qualitatively similar for all these cases.  Shown in
Fig.~\ref{Fig:Numerics} is ${\cal C}(\theta)$ for a 
half metal with $L_x = L_y = L_z$. 
The top dashed line in Fig.~\ref{Fig:Numerics} is the
variance of the conductance in the absence of spin-orbit
scattering. Without spin-orbit scattering, there is no angle-dependent
mesoscopic correction to the conductance, so ${\cal C}(\theta)$ is
independent of $\theta$.  For $\tau_{\up\parallel} \ll 1/E_{\up}$, conductance fluctuations
saturate at half their value without spin-orbit scattering. Changing
the magnetization by a small angle $\theta_{\rm c}$ changes the
mesoscopic conductance correction enough to lose all conductance
correlations. Our calculation shows
\begin{equation}
  \theta_{\rm c}
  \sim (\tau_{\up\parallel} E_{\up})^{1/2}
  \sim (\tau_{\up\parallel}/\tau_{\up})^{1/2} l/L,
\end{equation}
where $l$ is the mean free path.  In a realistic ferromagnet, the quantitative form of ${\cal
C}(\theta)$ is different, although the qualitative picture,
including the estimate for the correlation angle $\theta_{\rm c}$ is
the same as for the half metal (see Eq.~\ref{eq:main}).

Let us estimate the correlation angle $\theta_{\rm c}$
for the spin-orbit induced mesoscopic conductance fluctuations.
For the
highly disordered ferromagnetic wires used in the experiments of 
Refs.\ \onlinecite{kn:lee2004,kn:wei2005}, the mean free path $l$ 
is of the order of a few nm. Taking the spin-orbit times
$\tau_{\parallel}$ and $\tau_{\perp}$ within
an order of magnitude of the elastic scattering time $\tau$ (as is
appropriate for Co
\cite{kn:piraux1996}), we find $\theta_{\rm c} \sim (1 \times
10^{-8} \mbox{m})/L$. (Recall that $L$ has to be replaced by the phase
coherence length $L_{\phi}$ if $L_{\phi} < L$.) This would be
sufficiently small to explain the few conductance oscillations seen in
the experiment of Ref.\ \onlinecite{kn:wei2005}, for which 
$L_{\phi} \sim 30 \mbox{nm}$ and the conductance
was measured as a function of an external magnetic field that changed
the magnetization direction.

It is instructive to
compare the correlation angle $\theta_{\rm c}$ for spin-orbit induced
conductance fluctuations considered here to the correlation angle
arising from 
the coupling of the electron's charge to the internal magnetic field. 
The latter is $\sim \Phi_0/\Phi$, where $\Phi$ is 
the magnetic flux through the sample and $\Phi_0$ is the flux
quantum. Taking the internal magnetic field to be $\sim 2$T, as is
appropriate for Co, one finds
a correlation angle $\sim (2 \times 10^{-15} \mbox{m}^2)/L^2$. Hence, 
with the parameters taken above, the orbital effect will dominate for 
samples with size $L \gtrsim 2 \times 10^{-7}$m. This is in agreement
with Ref.\ \onlinecite{kn:wei2005}, where it was
shown that the orbital effect alone cannot account for the observed
conductance fluctuations \cite{kn:wei2005}. 

We thank I. Aleiner, D. Davidovi\'{c} and D. Ralph for discussions.
This work was supported in part by the NSF under grant no.\ DMR
0334499, by the Cornell Center for Nanoscale Systems under NSF 
grant no.\ EEC-0117770, and by the Packard foundation.

\end{document}